\documentclass[prl,twocolumn,floatfix,nofootinbib,amsmath,amssymb,floatfix]{revtex4} 
\usepackage{graphicx,color,dcolumn,booktabs,bm,mathrsfs}
\usepackage{longtable,lscape}
\usepackage{txfonts}
\usepackage{overpic}
\usepackage{amssymb}
\usepackage{indentfirst}
\usepackage{feynmf}   
\usepackage{slashed}  
\usepackage{cases}
\usepackage{color}
\usepackage{multirow}
\usepackage{epstopdf}
\usepackage{CJK}

\usepackage[colorlinks,
            citecolor=blue,
            anchorcolor=red,
            menucolor=red,
            linkcolor=red,
            filecolor=red,
            runcolor=red,
            urlcolor=blue,
            frenchlinks=red]{hyperref}
\usepackage{subfigure}
\usepackage{xcolor}

\newcommand{\be}{\begin{equation}}
\newcommand{\ee}{\end{equation}}
\newcommand{\bea}{\begin{eqnarray}} 
\newcommand{\eea}{\end{eqnarray}}

\begin{document}

\title{Precise determination of the $\eta \Lambda$ scattering length and effective range and relationship to the $\Lambda(1670)$ resonance }

\author{Man-Yu Duan $^{1,2,3}$}\email{duanmy@seu.edu.cn}
\author{Melahat Bayar $^{4}$}\email{melahat.bayar@kocaeli.edu.tr}
\author{Eulogio Oset$^{1,2,5}$}\email{oset@ific.uv.es}

\affiliation{$^1$Departamento de Física Teórica and IFIC, Centro Mixto Universidad de Valencia-CSIC Institutos de Investigación de Paterna, 46071 Valencia, Spain\\
$^2$School of Physics, Zhengzhou University, Zhengzhou 450001, China\\
$^3$School of Physics, Southeast University, Nanjing 210094, China\\
$^4$Department of Physics, Kocaeli University, 41380, Izmit, Turkey\\
$^5$Department of Physics, Guangxi Normal University, Guilin 541004, China}

\begin{abstract}
We use the Belle data on the $K^- p$ mass distribution of the $\Lambda_c^+ \to p K^- \pi^+$ reaction near the $\eta \Lambda$ threshold to determine the $\eta \Lambda $ scattering length and effective range. We show that from these data alone we can determine the value of $a$ with better precision than so far determined, and the value of $r_0$ for the first time. The addition of the $K^- p \to \eta \Lambda$ data allows us to improve the precision of these magnitudes, with errors smaller than $15\%$. We also determine with high precision the pole position of the $\Lambda(1670)$.
\end{abstract}

\date{\today}
\maketitle

\text{\textit{Introduction.}}— The issue of the $\eta \Lambda$ scattering length and the $\Lambda(1670)$ resonance are intimately tied and have had attention for some time. In Ref. \cite{chiang} the data for the $K^- p \to \eta \Lambda$ \cite{Berley:1965zz,ParticleDataGroup:1998hll} were analyzed, assuming the $\Lambda(1670)$ resonance to be the reason for the $\eta \Lambda$ interaction, and the $\eta \Lambda$ scattering length was obtained between $-0.21-i 1.02$ fm and $-0.42-i 0.44$ fm \footnote{We use here the notation $f^{-1} = - \frac{1}{a} + \frac{1}{2} r_0 k^2 -ik$, hence $a$ has opposite sign to Refs. \cite{chiang,nefkens}. We adapt all the result to our notation.}. A reanalysis of these data, also in connection to the $\Lambda(1670)$, was done in Ref. \cite{nefkens} with the results $a_{\eta \Lambda}= -0.64 (0.29) -i 0.80 (0.30)$. These results are barely compatible within the large errors, but in none of the papers an evaluation of the effective range was done. The advent of the chiral unitary approach for meson baryon interaction \cite{kaiser,angels,ollerulf,juan} allowed to put the $\Lambda(1670)$ in a broader context \cite{bennhold,juan,josean}, as a consequence of the interaction of $\bar{K} N$ with its coupled channels, which also gives rise to the two $\Lambda(1405)$ resonances \cite{ollerulf,cola,Cieply:2016jby} (see reviews on the subject in Refs. \cite{Oller:2000ma,Hyodo:2011ur,Meissner:2020khl}).

A fit to the $K^- p \to \eta \Lambda$ data using an empirical model which also includes the $\Lambda(1670)$ is done in Ref. \cite{xiejujun}, but no determination of the $\eta \Lambda$ scattering parameters is done. The $\Lambda(1670)$ is also shown to play an important role on the $\gamma p \to K^+ \eta \Lambda $ reaction near threshold \cite{xiezhong}, but once again no determination of the $\eta \Lambda$ scattering parameters is done. 
The authors of Ref. \cite{manyu} study the $\Lambda(1670)$ in the Cabibbo-favored process $\Lambda^+_c \to p K^- \pi^+$ decay, measured by the LHCb and Belle collaborations 
\cite{LHCb:2022sck,Belle:2022cbs}, and they associate the peak seen in the $K^- p$ invariant mass around $1670$ MeV to the $\Lambda(1670)$ resonance of molecular nature. 

 The Belle data \cite{Belle:2022cbs} show an unprecedented high resolution cusp like peak in the $K^- p$ invariant mass at the $\eta \Lambda$ threshold, which deserves special attention, which we provide in this paper. The $\Lambda^+_c \to p K^- \pi^+$ decay has been the object of theoretical study before \cite{Liu:2019dqc,BESIII:2015bjk,Filaseta:1987rx,Miyahara:2015cja,Ahn:2019rdr}. In Ref. \cite{Liu:2019dqc} a triangle mechanism for the production, that will be discussed below,  is assumed, and a cusp like in the $K^- \pi^+$ mass distribution associated with the $\Lambda(1670)$ is obtained at the $\eta \Lambda$ threshold. The same mechanism is used in Ref. \cite{manyu}. In Refs. \cite{BESIII:2015bjk,Filaseta:1987rx} that reaction is analyzed experimentally together with other reactions. The theoretical work of Ref. \cite{Miyahara:2015cja} deals with this reaction, together with other $\Lambda^+_c \to M B  \pi^+$ reactions, and predicts clear signals for the  $\Lambda(1670)$ for the $MB$ being $K^- p$ and $\eta \Lambda$. In Ref. \cite{Ahn:2019rdr} an effective Lagrangian approach to the reaction is used and again a clear signal is predicted for the $\Lambda(1670)$ around the $\eta \Lambda$ threshold. The related $\Lambda^+_c \to \pi^+ \eta \Lambda$ is also investigated in Ref. \cite{xielambc} along the same lines as in Ref. \cite{Miyahara:2015cja} and predictions are made for mass distributions related to the $\Lambda(1670)$. Also, in Ref. \cite{Lyu:2024qgc}, this latter reaction is used to investigate the existence of a $\Sigma^*(1/2^-)$ around the $\bar{K}N$ threshold and in Ref. \cite{Wang:2022nac} the same reaction is used to investigate this $\Sigma^*$ state together with the $\Lambda(1670)$ and the $a_0(980)$. Yet, with so much attention given to the $\Lambda^+_c \to p K^- \pi^+$ reaction and related $\Lambda^+_c$ decays, there were no attempts to obtain the $\eta \Lambda$ scattering length and effective range. The high resolution cusp like peak from the Belle experiment \cite{Belle:2022cbs} offers an opportunity that we exploit in the present work. The use of cusp data to determine scattering parameters has a long history and a time honored case is the determination of the $\pi \pi$ scattering length from the $K$ decay into pions \cite{Batley:2009ubw,Colangelo:2006va}. A recent review on this issue can be seen in Ref. \cite{Guo:2019twa}. 
 
In the present work we shall show that from the Belle data we can extract the $\eta \Lambda$ scattering length with a better precision than so far determined in Refs. \cite{chiang,nefkens}, and for the first time we can deduce the value of the effective range.  On the other hand we will combine the data for $K^- p \to \eta \Lambda$, measured with more precision in Ref. \cite{starostin}, with those of the $\Lambda^+_c \to p K^- \pi^+$ reaction, from where we will determine the $\eta \Lambda$ scattering length and the effective range with a precision of less than $15\%$.

\begin{figure}[tb]
\centering
\includegraphics[scale=0.7]{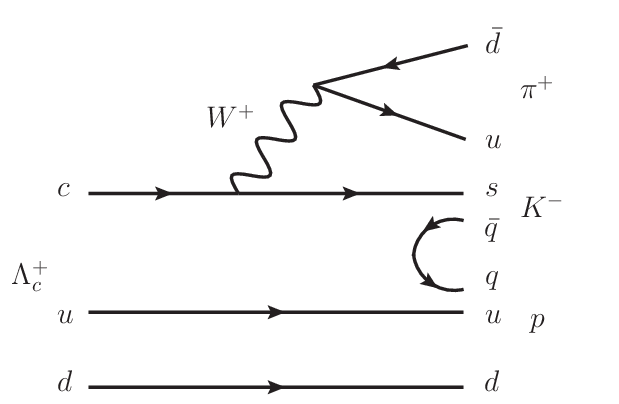}
\caption{Diagrammatic mechanism for $\Lambda_c^+ \to \pi^+ K^- p$ at the quark level with hadronization of the $su$ or $sd$ pairs.}
\label{fig1}
\end{figure}

\text{\textit{Formalism.}}— In order to establish the mechanism for the $\Lambda_c^+ \to p K^- \pi^+$ decay we start by looking at the original process at the quark level. We follow closely the formalism developed in Ref. \cite{Miyahara:2015cja} introducing external emission, which is depicted in Fig. \ref{fig1}. The $sud$ final quarks are hadronized into meson-baryon producing a $q\bar{q}$ pair, and using the $\Lambda_c$ wave function of Refs. \cite{Capstick:1986ter,Roberts:2007ni}, we have
\begin{eqnarray}\label{eq:quark}
\frac{1}{2}c(ud-du) &\to& \pi^+ \frac{1}{\sqrt{2}}s(ud-du) \nonumber\\
&\to& \pi^+ \frac{1}{\sqrt{2}}\sum_i(s\bar{q}_i q_i u d- s\bar{q}_i q_i d u).
\end{eqnarray}
By using the $q\bar{q}$ matrix $P$ in terms of pseudoscalar mesons, with the $\eta$, $\eta'$ mixing of Ref. \cite{Bramon:1992kr}, and ignoring the $\eta'$ which does not play a role in the reaction,
\begin{eqnarray}
    P &=& \left(\begin{array}{ccc}
        \frac{\pi^0}{\sqrt{2}}+\frac{\eta}{\sqrt{3}} & \pi^{+} & K^{+} \\
    \pi^{-} & -\frac{\pi^0}{\sqrt{2}}+\frac{\eta}{\sqrt{3}} & K^0  \\
    K^{-} & \bar{K}^0 & -\frac{\eta}{\sqrt{3}} \\
    \end{array}\right), 
\end{eqnarray}
Eq. \eqref{eq:quark} can be written as
\begin{eqnarray}
\frac{1}{2}c(ud-du) \to &&\pi^+ \frac{1}{\sqrt{2}} \Big( K^- u (ud-du) +\bar{K}^0 d (ud-du) \nonumber\\
 &&-\frac{\eta}{\sqrt{3}}s(ud-du) \Big) \,,
\end{eqnarray}
which using the wave functions of the octet of baryon in terms of quarks consistent with the formalism that we follow \footnote{The $\Lambda$ has opposite sign in the book of Close \cite{Close:1979bt}.} \cite{miya} can be written as 
\begin{equation}\label{eq:finalstate}
\pi^+ \left( \frac{1}{\sqrt{2}} K^- p + \frac{1}{\sqrt{2}} \bar{K}^0 n + \frac{1}{3}\eta \Lambda \right) \,,
\end{equation}
where we have taken into account that the $\Lambda_c$ has the spin wave function $\chi_{MA}$ and the baryon of the octet are written as $\frac{1}{\sqrt{2}}(\phi_{MS}\chi_{MS}+\phi_{MA}\chi_{MA})$.

\begin{figure}[tb]
\centering
\includegraphics[scale=0.7]{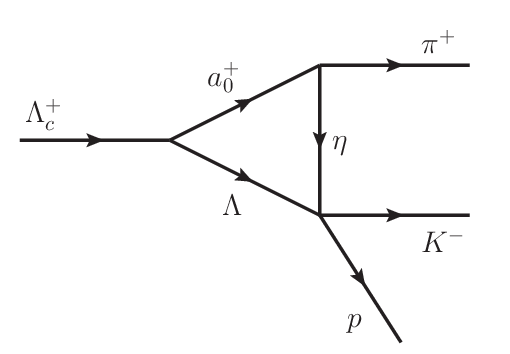}
\caption{Triangle mechanism assumed in Refs. \cite{Liu:2019dqc,manyu}.}
\label{fig2}
\end{figure}

At this point we must mention the mechanism used in Refs. \cite{Liu:2019dqc,manyu} depicted in Fig. \ref{fig2}. The $\Lambda_c \rightarrow a_0 \Lambda$ can be produced by the same mechanism as in Fig. \ref{fig1} hadronizing the $\bar{d}u$ component instead of the $su$ or $sd$. This hadronization gives rise to $u\bar{d} \to \sum_i u \bar{q}_iq_i\bar{d} \to (P^2)_{12}$, which leads to
\begin{equation}\label{eq:ts}
(\frac{\pi^0}{\sqrt{2}}+\frac{\eta}{\sqrt{3}})\pi^+ + \pi^+(-\frac{\pi^0}{\sqrt{2}}+\frac{\eta}{\sqrt{3}}) + K^+\bar{K}^0.
\end{equation}
Both the $\eta \pi^+ $ and the $K^+\bar{K}^0$ can give rise to the $a_0^+$ state \cite{npa}. However, while it looks like the $\eta \pi^+$ and $\pi^+ \eta$ terms in Eq. \eqref{eq:ts} add, this is not the case, because the order matters. Indeed, the vertex $W^{\mu} P_1 P_2$ goes as \cite{Gasser:1983yg,Scherer:2002tk}, $W^{\mu} [P, \partial_{\mu} P]$, which since the lower $cWs$ vertex $\gamma^{\mu}(1-\gamma_5)$ is dominated by the $\mu = 0$ component, gives rise to the structure $P^0_{\pi^+}-P^0_{\eta}$, for the $\pi^+ \eta$ term and $P^0_{\eta}-P^0_{\pi^+}$ for the $\eta \pi^+$ term, which cancel. Similarly, the $P^0_{K^+}-P^0_{\bar{K}^0}$ term coming from the $K^+\bar{K}^0$ component of Eq. \eqref{eq:ts} also largely cancels and the term becomes negligible. This discussion is substantiated by the extremely weak signal of $\Lambda_c^+ \to a_0^+(980) \Lambda$ observed in the $\Lambda_c^+ \to \pi^+ \eta \Lambda$ reaction of Ref. \cite{Belle:2020xku} (see diagonal in Fig. 5 of that reference). In addition the triangle diagram of Fig. \ref{fig2} is far away from developing a triangle singularity, as can be seen by application of Eq.(18) of Ref. \cite{triangle}. Thus, we safely neglect the mechanism of Fig. \ref{fig2} and stick to the external emission of Fig. \ref{fig1}, also assumed in Ref. \cite{Miyahara:2015cja} and in related $\Lambda_b$ decays \cite{Roca:2015tea}.

\begin{figure}[tb]
\centering
\includegraphics[scale=0.5]{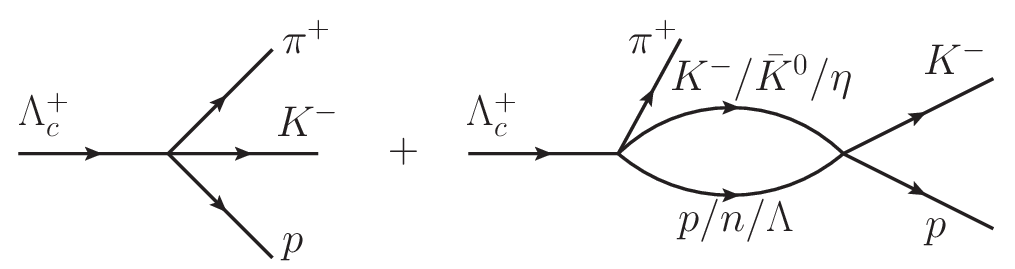}
\caption{Mechanism for $\Lambda_c^+ \to \pi^+ K^- p$ including meson-baryon rescattering.}
\label{fig3}
\end{figure}

In Eq. \eqref{eq:finalstate} we already have the $\pi^+ K^- p$ channel as final state of the tree level, but in order to generate the $\Lambda(1670)$ state we need to consider the meson-baryon rescattering as shown in Fig. \ref{fig3}.
The scattering matrix for the mechanism of Fig. \ref{fig3} is given by
\begin{eqnarray}\label{eq:t}
t=&&C\,\Big[ h_{K^-p}+ h_{K^-p}G_{K^-p}(M_\mathrm{inv})t_{K^-p,K^-p}(M_\mathrm{inv})  \nonumber\\
&&+ \, h_{\bar{K}^0 n}G_{\bar{K}^0 n}(M_\mathrm{inv})t_{\bar{K}^0 n,K^-p}(M_\mathrm{inv}) \nonumber\\
&&+ \, h_{\eta\Lambda}G_{\eta\Lambda}(M_\mathrm{inv})t_{\eta\Lambda,K^-p}(M_\mathrm{inv}) \Big] \,,
\end{eqnarray}
with $h_{K^-p}=h_{\bar{K}^0 n}=\frac{1}{\sqrt{2}}$, $h_{\eta\Lambda}=\frac{1}{3}$, and $C$ a global normalization constant, where $M_\mathrm{inv}$ is the $K^-p$ invariant mass, $G_i$ the meson baryon loop function regularized with a cut off, $q_{\mathrm{max}}$, as in Ref. \cite{angels} with $q_{\mathrm{max}}=630$ MeV, and $t_{ij}$ are the transition matrices between the 10 coupled channels used in Ref. \cite{angels}, $K^- p$, $\bar{K}^0 n$, $\pi^0 \Lambda$, $\pi^0 \Sigma^0$, $\eta \Lambda$, $\eta \Sigma^0$, $\pi^+ \Sigma^-$, $\pi^- \Sigma^+$, $K^+ \Xi^-$, $K^0 \Xi^0$. These matrices are obtained using the Bethe-Salpeter equation in coupled channels,
\begin{equation}\label{eq:BS}
  T=[1-VG]^{-1} \, V,
\end{equation}
with $V_{ij}$ the transition potential obtained in Ref. \cite{angels}
\begin{equation}\label{eq:Vij}
V_{ij}=- C_{ij} \dfrac{1}{4f^2} (k^0+k^{\prime 0});\, f=1.15f_{\pi},\, f_{\pi}=93\, \mathrm{MeV}, 
\end{equation}
with $k^0$, $k^{\prime 0}$ the initial and final meson energies in the meson-baryon center of mass frame and $C_{ij}$ coefficients obtained from the chiral Lagrangians which are given in Table 1 of Ref. \cite{angels}.

While in Ref. \cite{bennhold} the $\Lambda (1405)$ and $\Lambda (1670)$ states were simultaneously obtained, in terms of the formalism described above, here in order to obtain an exact reproduction of the experimental data of Ref. \cite{Belle:2022cbs} we will relax the potential by substituting $f^2$ in Eq. \eqref{eq:Vij} by $f_if_j$ with $f_i$ in a certain range. We also will accept some flexibility in the value of $q_{\mathrm{max}}$.

The $K^- p$ invariant mass distribution is obtained by means of 
\begin{equation}\label{eq:amplam}
\frac{d\Gamma}{dM_\mathrm{inv}}=\frac{1}{(2\pi)^3}\frac{1}{4M_{\Lambda_c}^2}p_{\pi^+}\tilde{p}_{K^-} |t|^2 \,,
\end{equation}
with $t$ given by Eq. \eqref{eq:t}, where the spin independence of the $t$ matrix has been assumed, given the $\gamma^0 \approx 1$ structure of the $cWs$ vertex, appearing in the transitions studied. The momenta $p_{\pi^+}$ and $\tilde{p}_{K^-}$ are the $\pi^+$ momentum in the $\Lambda_c$ rest frame and the $K^-$ momentum in the $K^- p$ rest frame, respectively. In addition, we consider a background term given by
\begin{equation}
\frac{d\Gamma_B}{dM_\mathrm{inv}}=\alpha-\beta M_\mathrm{inv} \,,
\end{equation}
which is added to Eq. \eqref{eq:amplam} and adjusted to the data.

The cross section for $K^-p \to \eta \Lambda$ is given by 
\begin{equation}
\sigma=\frac{1}{4\pi} \frac{M_{\Lambda}M_p}{s} \frac{q_{\eta}}{q_{K^-}} |t_{K^-p,\eta\Lambda}|^2 \,,
\end{equation}
where $q_{\eta}, q_{K^-}$ are the $\eta$ and $K^-$ momenta in the reaction.

In our approach, the scattering length and effective range for $\eta \Lambda$ are calculated from the $t_{\eta \Lambda, \eta \Lambda}$ amplitude taking into account the different normalization of our $t$ matrix and $f^{QM}(q)$ of Quantum mechanics.
\begin{equation}\label{eq:Tqm}
t=-\dfrac{8\pi \,\sqrt{s}}{2\, M_{\Lambda}}\; f^{\rm QM} \simeq -\dfrac{8\pi \,\sqrt{s}}{2\, M_{\Lambda}}\; \dfrac{1}{-\frac{1}{a}+\frac{1}{2}r_0 k^2 -ik},
\end{equation}
from where one gets \cite{Li:2023pjx}
\begin{equation}\label{eq:aj2}
-\dfrac{1}{a}=\left. -\dfrac{8\pi \,\sqrt{s}}{2\, M_{\Lambda}}\; (t_{\eta\Lambda,\eta\Lambda})^{-1}\right|_{\sqrt{s}_{\rm th}},
\end{equation}
\begin{eqnarray}\label{eq:r0j}
r_{0}=\left. \dfrac{1}{\mu}\; \left[ \dfrac{\partial}{\partial \sqrt{s}}\,
\left( -\dfrac{8\pi \,\sqrt{s}}{2\, M_{\Lambda}}\; (t_{\eta\Lambda,\eta\Lambda})^{-1}+ik \right)\right]\right|_{\sqrt{s}_{\rm th}},
\end{eqnarray}
with ${\sqrt{s}_{\rm th}}=m_{\Lambda}+m_{\eta}$ and $k$ the $\eta$ momentum in the reaction.

\begin{figure*}[t]
\centering
\includegraphics[scale=0.7]{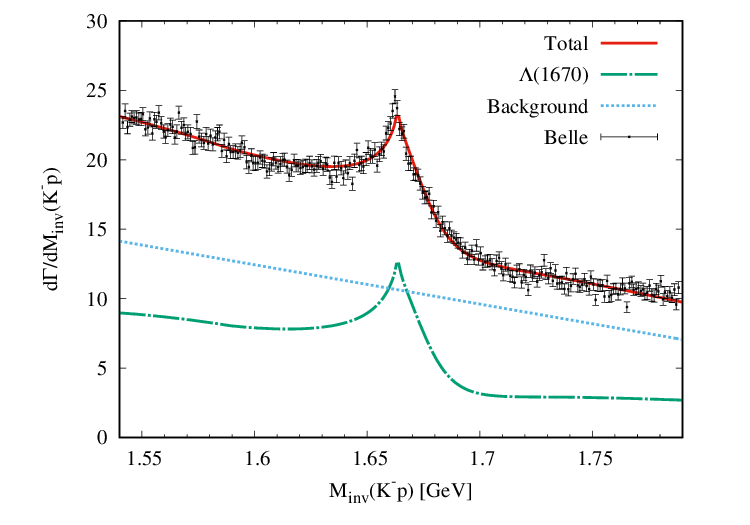}\put(-210,160){(a)}
\includegraphics[scale=0.7]{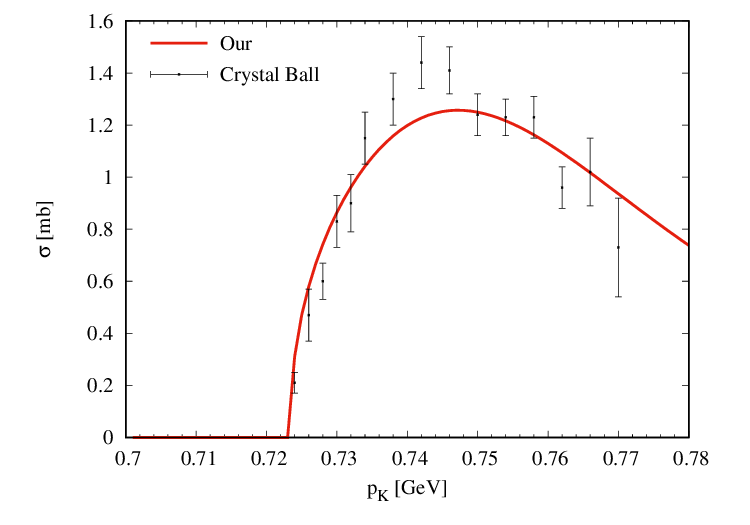}\put(-35,160){(b)}
\caption{Results of the fits to the $\Lambda_c^+ \to \pi^+ K^- p$ (a) \cite{Belle:2022cbs} and $K^-p \to \eta\Lambda$ (b) \cite{starostin} data.}
\label{fig4}
\end{figure*}

\text{\textit{Results.}}—We conduct several fits to the data of Ref. \cite{Belle:2022cbs} using different ranges of data,
\begin{eqnarray}
(\romannumeral 1) &&1540 < \sqrt{s} < 1790 \; \mathrm{MeV},  \nonumber\\
(\romannumeral 2) &&1600 < \sqrt{s} < 1720 \; \mathrm{MeV},  \nonumber\\
(\romannumeral 3) &&1640 < \sqrt{s} < 1685 \; \mathrm{MeV}.
\end{eqnarray}
We conduct fits varying $f_i$ and $q_{\mathrm{max},i}$. However, in order to conserve isospin in the potential (it is a bit violated in the $t_{ij}$ amplitudes due to mass differences of the particles in the same isospin multiplets) we impose
\begin{eqnarray}
&&f_{K^- p}=f_{\bar{K}^0 n}, f_{\pi^0\Sigma^0}=f_{\pi^+\Sigma^-}=f_{\pi^-\Sigma^+}, f_{K^+ \Xi^-}=f_{K^0 \Xi^0},\, \mathrm{and} \nonumber\\
&&q_{\mathrm{max},K^- p}=q_{\mathrm{max},\bar{K}^0 n}, q_{\mathrm{max},\pi^0\Sigma^0}=q_{\mathrm{max},\pi^+\Sigma^-}=q_{\mathrm{max},\pi^-\Sigma^+},   \nonumber\\
&&q_{\mathrm{max},K^+ \Xi^-}=q_{\mathrm{max},K^0 \Xi^0}.
\end{eqnarray}
We allow $f_i$ in the range $40-180$ MeV and $q_{\mathrm{max}}$ in the range $400-1500$ MeV. In addition we have the global $C$ constant and the two parameters of the background, $\alpha$, $\beta$. In total we have $15$ free parameters. Obviously, there are correlations between these parameters, in particular between $f_i$ and $q_{\mathrm{max},i}$. 

A common, and accurate, method using fits to determine the values of the observables and their uncertainties is the use of the resampling (bootstrap) method \cite{Press:1992zz,Efron:1986hys,Albaladejo:2016hae}. Random center points of data are generated with a Gaussian distribution and a fit to these new points with the same experimental errors is conducted. New random data sets are generated and a best fit is conducted in each case. With the parameters obtained in each fit, the values of the observables are evaluated. After a large number of runs, of the order of 100, the average and the dispersion for all these observables are calculated.

The results that we get from these fits to the Belle data are
\begin{eqnarray}\label{eq:onlybelle}
a&=&-(0.70\pm 0.16)-i(0.75\pm 0.24),  \nonumber\\
r_0&=&-(4.98\pm 0.96)+i(1.05\pm 0.35),
\end{eqnarray}
the results for $a$ are compatible with those obtained in Refs. \cite{chiang,nefkens}, but the errors are smaller, of the order of $25-30\%$, one half of those determined before. The novelty is that for the first time we are able to determine the effective range, with an uncertainty of about $20\%$ for the real part and $30\%$ for the imaginary part.

We conduct now a second type of fits, where we also include the $K^- p \rightarrow \eta \Lambda$ data of Ref. \cite{starostin}. We find now
\begin{eqnarray}\label{eq:global}
a&=&-(0.58\pm 0.09)-i(0.59\pm 0.04),  \nonumber\\
r_0&=&-(6.68\pm 0.93)+i(2.10\pm 0.34),
\end{eqnarray}
the results are compatible with those of Eq.\eqref{eq:onlybelle}, within errors, but the errors are now much smaller, of the order of smaller than $15\%$. This is a high precision determination, in particular for $r_0$, which is determined here for the first time.

We have conducted another sort of fits, this time to the $K^- p \rightarrow \eta \Lambda$ data alone. The results obtained are illustrative. We obtain
\begin{eqnarray}\label{eq:onlycrystal}
a&=&-(0.71\pm 0.10)-i(0.46\pm 0.04),  \nonumber\\
r_0&=&-(8.71\pm 0.76)+i(1.58\pm 0.33),
\end{eqnarray}
the errors in Re$(a)$ are of the order of $15\%$, but in Im$(r_0)$ they are of the order of $20\%$, hence, these errors are a bit bigger than those from the cusp in $\Lambda_c^+ \rightarrow p K^- \pi^+$ to determine the scattering length and effective range. Yet, in the fit of Eq. \eqref{eq:global} a large range of energies is fitted. We take as our final results those of Eq. \eqref{eq:global} that fit all data.

Finally, in order to visualize the quality of the fits that we carry, we take one of the fits that leads to the values of Eq. \eqref{eq:global}, where all data are considered, and show the result in Fig. \ref{fig4}. We can see that the quality of the fit is excellent.

We also look for poles, in the second Riemann sheet, setting
\begin{equation}
G^{II}(\sqrt{s})=G^{I}(\sqrt{s})+i \frac{M_B}{2\pi\sqrt{s}}q_{\mathrm{on}},
\end{equation}
for $\mathrm{Re}(\sqrt{s})>M+m$, $q_{\mathrm{on}}=\lambda^{1/2}(s, m^2, M^2)/(2\sqrt{s})$,$(\mathrm{Im}(q_{\mathrm{on}})>0)$, with $M$, $m$ the baryon and meson mass of the channel, and we find a pole around $1670$ MeV of isospin $I=0$. We determine the uncertainties as before and find
\begin{equation}
M_{\Lambda(1670)}-i\frac{\Gamma_{\Lambda(1670)}}{2}=( 1669.40 \pm 1.06 )-i( 21.46 \pm 1.76 ),
\end{equation}
which appears around $6$ MeV above the $\eta\Lambda$ threshold. These results are in line with those determined in Ref. \cite{Sarantsev:2019xxm}, $(1676 \pm 2)-i (16.5 \pm 2)$ MeV, and Ref. \cite{Kamano:2015hxa} $1669^{+3}_{-8}-i (9.5^{+9}_{-1})$ MeV, obtained from $\bar{K}N$ multichannel analysis, but a bit more precise due to the new Belle data.

\text{\textit{Conclusions.}}—We have taken the results of the high precision Belle data on the $\Lambda^+_c \to p K^- \pi^+$ reaction, with a neat cusp for the $K^- p$ mass distribution around the $\eta \Lambda$ threshold, and conducted a fit to these data. We have allowed some flexibility in the parameters of the chiral Lagrangians for the potentials and the regulators of the loops, and from many fits using the resampling technique we have determined the values of the scattering length and effective range. We obtain values for the scattering length that improve the accuracy with respect to former determinations based on the $K^- p \to \eta \Lambda$ data, and determine the effective range for the first time. We have also determined the pole position of the $\Lambda(1670)$, which is compatible with other determinations, but also improve in the accuracy.

\text{\textit{Acknowledgments.}}—We thank En Wang for useful comments and information. This work is partly supported by the Spanish Ministerio de Economia y Competitividad (MINECO) and European FEDER funds under Contracts No. FIS2017-84038-C2-1-P B, PID2020-112777GB-I00, and by Generalitat Valenciana under contract PROMETEO/2020/023. This project has received funding from the European Union Horizon 2020 research and innovation programme under the program H2020-INFRAIA-2018-1, grant agreement No. 824093 of the STRONG-2020 project.



\end{document}